\documentclass[prd,twocolumn,showpacs,floatfix,amsmath,nofootinbib,amssymb,floatfix]{revtex4}
\usepackage{graphicx,color,dcolumn,booktabs,bm,multirow}
\usepackage{longtable,lscape}
\usepackage{txfonts}
 \usepackage{enumitem}
\usepackage{overpic}
\usepackage{amssymb}
\usepackage{indentfirst}
\usepackage{feynmf}   
\usepackage{slashed}  
\usepackage{cases}
\usepackage{color}
\usepackage{multirow}
\usepackage{epstopdf}
\usepackage{graphicx,color,dcolumn,booktabs,bm}

\usepackage[colorlinks,
            citecolor=blue,
            anchorcolor=red,
            menucolor=red,
            linkcolor=red,
            filecolor=red,
            runcolor=red,
            urlcolor=blue,
            frenchlinks=red]{hyperref}
\usepackage{amsmath}
\usepackage{epsfig}
\usepackage{graphicx}
\usepackage{txfonts}

\newcommand{\tccs}{T_{cc\bar{s}}}
\newcommand{\br}{\boldsymbol{r}}

\begin{document}

\title{Predictions of the Strange partner of $T_{cc}$ in the quark delocalization color screening model}

\author{Xuejie Liu$^1$}\email[E-mail: ]{1830592517@qq.com}
\author{Dianyong Chen$^{2,4}$\footnote{Corresponding author}}\email[E-mail:]{chendy@seu.edu.cn}
\author{Hongxia Huang$^3$}\email[E-mail:]{hxhuang@njnu.edu.cn}
\author{Jialun Ping$^3$}\email[E-mail: ]{jlping@njnu.edu.cn}
\affiliation{$^1$School of Physics, Henan Normal University, Xinxiang 453007, P. R. China}
\affiliation{$^2$School of Physics, Southeast University, Nanjing 210094, P. R. China}
\affiliation{$^3$Department of Physics, Nanjing Normal University, Nanjing 210023, P.R. China}
\affiliation{$^4$Lanzhou Center for Theoretical Physics, Lanzhou University, Lanzhou 730000, P. R. China}

\date{\today}

\begin{abstract}
Inspired by the detection of $T_{cc}$ tetraquark state by LHCb Collaboration, we preform a systemical investigation of the low-lying doubly heavy charm tetraquark states with strangeness in the quark delocalization color screening model in the present work. Two kinds of configurations, the meson-meson configuration and diquark-antidiquark configuration,  are considered in the calculation. Our estimations indicate that the coupled channel effects play important role in the multiquark system, and a bound state with $J^{P}=1^{+}$ and a resonance state with $J^{P}=0^{+}$ have been predicted. The mass of the bound state is evaluated to be $(3971\sim3975)$ MeV, while the mass and width of the resonance are determined to be $(4113\sim4114)$ MeV and $(14.3\sim 16.1)$ MeV, respectively.
\end{abstract}

\maketitle

\setcounter{totalnumber}{5}

\section{\label{sec:introduction}Introduction}

In the recent two decades, an increasing number of charmonium-like states have been observed experimentally, which provide a good opportunity of searching for multiquark states. As the first confirmed charmonium-like state, $Z_c(3900)$ was first observed in the year of 2013 by the BESIII\cite{BESIII:2013ris} and Belle~\cite{Belle:2013yex} Collaborations in the $\pi^{+}J/\psi$ invariant mass spectrum of the process $e^+e^- \to \pi^+\pi^- J/\psi$ at a center of mass energy of 4.26 GeV, and then the authors of Ref.~\cite{Xiao:2013iha} further confirmed the existence of $Z_c(3900)$ by using the data sample collected by CLEO-c detector in the same process but at $\sqrt{s}=4.170$ GeV. The partial wave analysis of the process $e^+e^-\to \pi^+\pi^- J/\psi$  with the data sample accumulated at $\sqrt{s}=4.23$ and $4.26$ GeV indicated that the spin and parity of the $Z_c(3900)^\pm$ state are $1^+$~\cite{BESIII:2017bua}. The observations indicate that such a new particle can not be simply interpreted in the conventional quark-antiquark and three-quark schemes. Thus, some exotic interpretations, such as tetraquark state~\cite{Agaev:2016dev,Dias:2013xfa,Wang:2013vex,Deng:2014gqa}, hadronic molecular state~\cite{Wang:2013daa,Wilbring:2013cha,Dong:2013iqa,Ke:2013gia,Gutsche:2014zda,Esposito:2014hsa,Chen:2015igx,Gong:2016hlt,Ke:2016owt}, have been proposed. Besides the resonance interpretations, $Z_c(3900)$ has also been considered as the kinematic effects~\cite{Swanson:2014tra,HALQCD:2016ofq,Szczepaniak:2015eza,Chen:2011xk,Chen:2013coa,Swanson:2015bsa}, which indicated that $Z_c(3900)$ was not a genuine resonance.

\begin{figure}[t]
\includegraphics[scale=0.5]{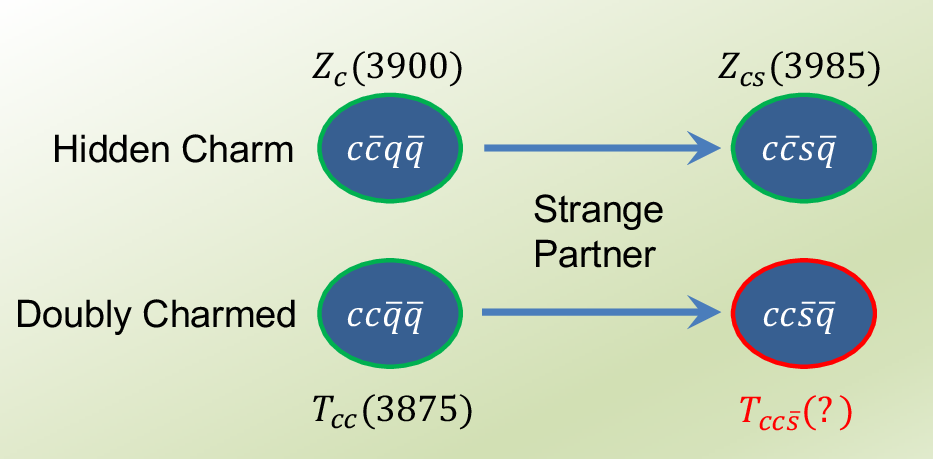}
\caption{(Color online). The similarity of the hidden charm and doubly charmed states. Hereinafter, $T_{cc\bar{s}}$ is used to refer the doubly charmed state with strangeness. }
\label{Fig:Analog}
\end{figure}

In the resonance frame, the quark component of $Z_c(3900)$ is $c\bar{c}q\bar{q}$. The flavor independence of the strong interactions naturally indicates the possible existence of the strange partner of $Z_{c}(3900)$, whose quark components are $c\bar{c}s \bar{q}$. Such kind of charmonium-like states with strangeness have been predicted theoretically in various model, such as tetraquark scenarios~\cite{Ebert:2008kb, Ferretti:2020ewe}, hadronic molecular model~\cite{Lee:2008uy, Dias:2013qga}, the hadro-quarkonium model~\cite{Ferretti:2020ewe} and initial single chiral particle emission mechanism~\cite{Chen:2013wca}. In the year of 2020, the BES III Collaboration observed a new states named $Z_{cs}(3985)$ in the $K^+$ recoil mass distributions of the process $e^{+}e^{-}\rightarrow K^{+}D_{s}^{-}D^{*0}/K^{+}D_{s}^{*-}D^{0}$~\cite{BESIII:2020qkh}. Later on, the LHCb Collaboration reported their observation of two exotic structures, $Z_{cs}(4000)$ and $Z_{cs}(4220)$, in the $J/\psi K^{+}$ invariant mass spectrum of the $B^{+}\rightarrow J/\psi \phi K^{+}$ decay in 2021~\cite{LHCb:2021uow}. Since the observed masses of $Z_{cs}(3985)$ and  $Z_{cs}(4000)$ were similar, these two states may be considered as the same one (hereinafter, we use $Z_{cs}(3985)$ to refer to this state). It's interesting to notice that $Z_c(3900)$ is located in the vicinity of the $D^\ast \bar{D}$ threshold, while $Z_{cs}(3985)$ is close to $D_s^\ast \bar{D}$ threshold, thus one can consider $Z_{cs}(3985)$ as a strange partner of $Z_c(3900)$. Consequently, the hadronic molecular~\cite{Chen:2021erj,Wan:2020oxt,Xu:2020evn,Albuquerque:2021tqd,Ozdem:2021yvo,Ozdem:2021hka,Wang:2020rcx,Meng:2020ihj,Yang:2020nrt,Du:2020vwb}, compact tetraquark ~\cite{Maiani:2021tri,Shi:2021jyr,Giron:2021sla} and hadro-quarkonium ~\cite{Ferretti:2020ewe} scenarios have been proposed to decode the nature of $Z_{cs}(3985)$.

 In the naive multiquark scenario, if there are multiquark states composed of $c\bar{c} q\bar{q}$, the states composed of $cc\bar{q}\bar{q}$ are also expected to exist and have been considered to be the molecular $D^{*+}D^{0}$ states~\cite{Chen:2021vhg,Santowsky:2021bhy,Chen:2021tnn,Feijoo:2021ppq,Wang:2021yld,Wang:2021ajy,Ren:2021dsi,Deng:2021gnb,Xin:2021wcr,Azizi:2021aib,Ozdem:2021hmk,Albaladejo:2021vln,Dai:2021vgf,Dai:2022ulk,Du:2021zzh,Zhao:2021cvg,Ke:2021rxd}, and compact states~\cite{Guo:2021yws,Weng:2021hje,Agaev:2021vur}. Recently, the LHCb Collaboration reported the observation of the first doubly charmed tetraquark state $T_{cc}^{+}(3875)$ in the $D^{0}D^{0}\pi^{+}$ mass spectrum just below the $D^{*+}D^{0}$ mass threshold~\cite{LHCb:2021vvq,LHCb:2021auc} with $I(J^{P})=1(1^{+})$. As indicated in Fig.~\ref{Fig:Analog}, the quark components of $T_{cc}(3875)$ are $cc\bar{q}\bar{q}$, which indicate that $T_{cc}(3875)$ could be a good candidate of compact tetraquark  state. In Refs.~\cite{Guo:2021yws,Weng:2021hje}, the authors investigated the mass spectrum of the $S-$wave doubly heavy tetraquark states $QQ\bar{q}\bar{q}$ based on the improved chromomagnetic interaction model and found a stable $cc\bar{u}\bar{d}$ tetraquark state with $I(J^{P})=0(1^{+})$ below the $D^{*+}D^{0}$ threshold, which is well consistent with the observed $T_{cc}^{+}(3875)$. Moreover, the QCD sum rule estimation in Ref.~\cite{Agaev:2021vur} also supported the compact tetraquark interpretation. In addition, the observed mass of $T_{cc}^+(3875)$ is only several hundred keV below the threshold of $D^0 D^{\ast +}$, which imply that $T_{cc}^+(3875)$ could be interpreted as a shallow molecular state composed of $D^0 D^{\ast+}+h.c.$. Further estimations by using the quark models~\cite{Chen:2021vhg,Santowsky:2021bhy,Chen:2021tnn,Feijoo:2021ppq,Wang:2021yld,Wang:2021ajy,Ren:2021dsi,Deng:2021gnb}, QCD sum rules~\cite{Xin:2021wcr,Azizi:2021aib,Ozdem:2021hmk}, heavy quark symmetry~\cite{Albaladejo:2021vln,Dai:2021vgf,Dai:2022ulk,Du:2021zzh} and Bethe-Salpeter equations~\cite{Zhao:2021cvg,Ke:2021rxd} indicated that $T_{cc}^+(3875)$ could be a good candidate of $D^0 D^{\ast+}+h.c.$ molecular state.

Similar to the relation between $Z_{cs}(3985)$ and $Z_c(3900)$, one can expect the existence of the strange partner of $T_{cc}(3875)$, i.e., the tetraquark states composed of $cc\bar{q}\bar{s}$. Actually, before the observation of $T_{cc}^{+}(3875)$, the Lattice QCD estimations in Ref.~\cite{Junnarkar:2018twb} predicted that the $T_{cc\bar{s}}$ state with $J^P=1^+$ was about 10 MeV below the threshold of $D^+ D_s^{\ast -}$, while the estimations by using the heavy quark symmetry in Ref.~\cite{Eichten:2017ffp} found its mass to be about 180 MeV above the corresponding threshold. In Ref.~\cite{Luo:2017eub}, the predicted $\tccs$ tetraquark state with $J^P=1^+$ was below the threshold of $D^+ D_s^{\ast -}$, while those with $J^P=0^+$ and $2^+$ were both above the corresponding thresholds. After the observation of $T_{cc}^{+}$, the authors in  Ref.~\cite{Dai:2021vgf} took advantage of the experimental information on the binding energy of $T_{cc}^{+}$ to fix the cutoff regulator of the loops in the Bethe-Salapeter equation and a  $D_{s}^{*}D^{*}$ bound state with $J^{P}=1^{+}$ was predicted. Besides, the color-magnetic model estimations in Ref.~\cite{Karliner:2021wju} implied that both $T_{cc}^{+}$ and $\tccs^{+}$ system could be stable against the strong interactions. However, the state $\tccs^{+}$ was not found in the quark model but if the mixing of S$-$D wave was taken into account, this state may be obtained~\cite{Deng:2021gnb}. As mentioned above, theorists have not reach an agreement on the existence of $\tccs$ tetraquark states. In the present work, we perform a system estimations of $\tccs$ system by using the quark delocalization color screening model (QDCSM) in an attempt to further explore the existence of the possible bounded and resonant states in the $\tccs$ system.

This paper is organized as follows. After the introduction, the details of the QDCSM and resonating group method (RGM) are presented in Section~\ref{model}. Our numerical results and the related discussions for $T_{cc\bar{s}}$ system are given in Section~\ref{results}, and the last section is devoted to a short summary.

\section{Quark delocalization color screening model and  the resonanting group method }{\label{model}}

\subsection{Quark delocalization color screening model}
The QDCSM is an extension of the native quark cluster model~\cite{DeRujula:1975qlm,Isgur:1978xj,Isgur:1978wd,Isgur:1979be} and also developed with aim of addressing multiquark systems. For the tetraquark system, the Hamiltonian reads,
\begin{equation}
H = \sum_{i=1}^{4} \left(m_i+\frac{\boldsymbol{p}_i^2}{2m_i}\right)-T_{CM}+\sum_{j>i=1}^4V(r_{ij}),\\
\end{equation}
where $T_{CM}$ is the center-of-mass kinetic energy, who is usually subtracted without losing generality since we mainly focus on the internal relative motions of the multiquark system. The interplay is two body potential, which includes color-confining potential $V_{\mathrm{CON}}$, one-gluon exchange potential $V_{\mathrm{OGE}}$, and the potential results from Goldstone-boson exchange, $V_{\chi}$, i.e.,
\begin{equation}
V(r_{ij}) = V_{\mathrm{CON}}(r_{ij})+V_{\mathrm{OGE}}(r_{ij})+V_{\chi}(r_{ij}).
\end{equation}

In the present work, we focus on the $S-$wave low-lying positive  $\tccs$ tetraquark system with positive parity. In this case, the spin-orbit and tensor interactions vanish and the potential $V_{\mathrm{OGE}}(r_{ij})$ becomes,
\begin{eqnarray}
\nonumber
V_{\mathrm{OGE}}(r_{ij}) &=& \frac{1}{4}\alpha_s^{q_{i}q_{j}} \boldsymbol{\lambda}^{c}_i \cdot
\boldsymbol{\lambda}^{c}_j \\
&&\left[\frac{1}{r_{ij}}-\frac{\pi}{2}\delta(\boldsymbol{r}_{ij})(\frac{1}{m^2_i}+\frac{1}{m^2_j}
+\frac{4\boldsymbol{\sigma}_i\cdot\boldsymbol{\sigma}_j}{3m_im_j})\right],
\end{eqnarray}
where $m_{i}$ is the quark mass,  $\boldsymbol{\sigma}_i$ and $\boldsymbol{\lambda^{c}}_i$ are the Pauli matrices and SU(3) color matrices, respectively. The $\alpha_s^{q_{i}q_{j}}$ is the quark-gluon coupling constant, which offers a consistent description of mesons from light to heavy-quark sector. The values of $\alpha_{ij}$ are associated with the quark flavors and in the present work they are fixed by reproducing the mass difference of the low-lying mesons with $S=0$ and $S=1$.

\begin{table*}[t]
\centering
\caption{Three sets of model parameters involved in the present estimations. \label{Tab:Para}}
\begin{tabular}{p{2.5cm}<\centering p{2.5cm}<\centering p{2.5cm}<\centering p{2.5cm}<\centering p{2.5cm}<\centering p{2cm}<\centering c}
\toprule[1pt]
 & Parameters & QDCSM1 & QDCSM2 & QDCSM3\\
\midrule[1pt]
           & $m_u$(MeV) & 313  & 313 & 313\\
Quark Mass & $m_s$(MeV) & 536  & 536 & 536\\
           & $m_c$(MeV) & 1728 & 1728& 1728\\
\midrule[1pt]
          & b(fm)      &0.29  &0.3  &0.315 \\
           &  $a_c$(MeV $fm^{-2}$) &101     &101      &101 \\
           &  $V_{0_{uu}}$(MeV) &-2.3928 &-2.2543  &-2.0689 \\
Confinement   &  $V_{0_{us}}$(MeV) &-1.9137 &-1.7984  &-1.6429 \\
           &  $V_{0_{uc}}$(MeV) &-1.4175 &-1.3231  &-1.2052 \\
           &  $V_{0_{ss}}$(MeV) &-1.3448 &-1.2826  &-1.2745 \\
           &  $V_{0_{sc}}$(MeV) &-0.7642 &-0.6739  &-0.5452 \\
           &  $V_{0_{cc}}$(MeV) &0.6063  &0.7555   &0.9829 \\
\midrule[1pt]
        &  $\alpha_{s}^{uu}$ &0.2292  &0.2567   &0.3019 \\
           &  $\alpha_{s}^{us}$ &0.2655  &0.2970   &0.3484 \\
           &  $\alpha_{s}^{uc}$ &0.3437  &0.3805   &0.4405 \\
   OGE        &  $\alpha_{s}^{ss}$ &0.3856  &0.3604   &0.3360 \\
           &  $\alpha_{s}^{sc}$ &0.5969  &0.6608   &0.7649 \\
           &  $\alpha_{s}^{cc}$ &1.5101  &1.6717   &1.9353 \\
\bottomrule[1pt]	
\end{tabular}
\end{table*}

The confining potential $V_{\mathrm{CON}}(r_{ij})$ is,
\begin{equation}
 V_{\mathrm{CON}}(r_{ij}) =  -a_{c}\boldsymbol{\lambda^{c}_{i}\cdot\lambda^{c}_{j}}\left[f(r_{ij})+V_{0_{q_{i}q_{j}}}\right],
\end{equation}
where the $V_{0_{q_{i}q_{j}}}$ is determined by the mass differences of the theoretical esmations and experimental measurement of each kind of meson, which is also quark flavor related parameter. In the QDCSM, the function $f(r_{ij})$ is defined as,
\begin{equation}
 f(r_{ij}) =  \left\{ \begin{array}{ll}r_{ij}^2  &\qquad \mbox{if }i,j\mbox{ occur in the same cluster} , \\
\frac{1 - e^{-\mu_{ij} r_{ij}^2} }{\mu_{ij}} & \qquad \mbox{if }i,j\mbox{ occur in different cluster} , \\
\end{array} \right.
\end{equation}
where the color screening parameter $\mu_{ij}$ relevant to the light quarks can be determined by fitting the deuteron properties, $NN$ and $NY$ scattering phase shifts~\cite{Chen:2011zzb,Ping:1993me,Wang:1998nk}, which are  $\mu_{qq}= 0.45$, $\mu_{qs}= 0.19$
and $\mu_{ss}= 0.08$. The parameter $\mu_{ij}$ satisfy the relation $\mu_{qs}^{2}=\mu_{qq}\mu_{ss}$, where $q$ represents $u$ or $d$ quark. When extending to the heavy-quark case, we found that the dependence of the parameter $\mu_{cc}$ is rather weak in the calculation of the spectrum of $P_{c}$ states by taking the value of $\mu_{cc}$ from $10^{-4}$ to $10^{-2}\ \mathrm{fm}^{-2}$~\cite{Huang:2015uda}. Moreover, when $\mu_{ij}$ is rather small, the exponential function can be approximated to be,
\begin{eqnarray}\label{muij}
  e^{-\mu_{ij}r_{ij}^{2}} &=& 1-\mu_{ij}r_{ij}^{2}+\mathcal{O}(\mu_{ij}^2 r_{ij}^4).
\end{eqnarray}
in the small $r$ region. Accordingly, the confinement potential between two clusters is approximated to be,
\begin{eqnarray}
  V_{\mathrm{CON}}(r_{ij}) &=&  -a_{c}\boldsymbol{\mathbf{\lambda}}^c_{i}\cdot
\boldsymbol{\mathbf{
\lambda}}^c_{j}~\left(\frac{1-e^{-\mu_{ij}\mathbf{r}_{ij}^2}}{\mu_{ij}}+ V_{0_{ij}}\right) \nonumber \\
  ~ &\approx & -a_{c}\boldsymbol{\mathbf{\lambda}}^c_{i}\cdot
\boldsymbol{\mathbf{ \lambda}}^c_{j}~(r_{ij}^2+ V_{0_{ij}}),
\end{eqnarray}
which is the same with the expression of two quarks in the same cluster. Thus, when the value of the $\mu_{ij}$ is very small, the screened confinement will return to the quadratic form, which is why the results are insensitive to the value of $\mu_{cc}$. So in the present work, we take $\mu_{cc}=0.01\ \mathrm{fm}^{-2}$. Then $\mu_{sc}$ and $\mu_{uc}$ are obtained by the relation $\mu^{2}_{sc}=\mu_{ss}\mu_{cc} $ and $\mu^{2}_{uc}=\mu_{uu}\mu_{cc}$, respectively.

The Goldstone-boson exchange interactions between light quarks appear because the dynamical breaking of chiral symmetry. For the $\tccs$ system, the $\pi$ exchange interaction vanishes because there is no unflavor quark pair in the tetraquark state, and then the concrete form of the Goldstone-boson exchange potential becomes,
\begin{eqnarray}
V^{\chi}_{ij}  &=& V_{K}(\boldsymbol{r}_{ij})\sum_{a=4}^7\lambda
_{i}^{a}\cdot \lambda _{j}^{a}+ \nonumber\\
&&V_{\eta}(\boldsymbol{r}_{ij})\left[\left(\lambda _{i}^{8}\cdot
\lambda _{j}^{8}\right)\cos\theta_P-(\lambda _{i}^{0}\cdot
\lambda_{j}^{0}) \sin\theta_P\right], \label{sala-Vchi1}
\end{eqnarray}
with
\begin{eqnarray}
\nonumber
V_{\chi}(\boldsymbol{r}_{ij}) &=&  {\frac{g_{ch}^{2}}{{4\pi}}}{\frac{m_{\chi}^{2}}{{\
12m_{i}m_{j}}}}{\frac{\Lambda _{\chi}^{2}}{{\Lambda _{\chi}^{2}-m_{\chi}^{2}}}}
m_{\chi}                                 \\
&&\left\{(\boldsymbol{\sigma}_{i}\cdot\boldsymbol{\sigma}_{j})
\left[ Y(m_{\chi}\,r_{ij})-{\frac{\Lambda_{\chi}^{3}}{m_{\chi}^{3}}}
Y(\Lambda _{\chi}\,r_{ij})\right] \right\},\nonumber \\
&& ~~~~~~\chi=\{K, \eta\},
\end{eqnarray}
where $Y(x)=e^{-x}/x$ is the standard Yukawa function. The $\boldsymbol{\lambda^{a}}$ is the SU(3) flavor Gell-Mann matrix. The mass of the $K$ and $\eta$ meson is taken from the experimental value~\cite{ParticleDataGroup:2018ovx}. The chiral coupling constant, $g_{ch}$, is determined from the $\pi NN$ coupling constant through,
\begin{equation}
\frac{g_{ch}^{2}}{4\pi}=\left(\frac{3}{5}\right)^{2} \frac{g_{\pi NN}^{2}}{4\pi} {\frac{m_{u,d}^{2}}{m_{N}^{2}}},
\end{equation}
where the SU(3) flavor symmetry only broken by the different masses of the light quarks. All the other model parameters are  the same as the ones in Ref.~\cite{Liu:2021xje}, where three different sets of parameters were used to study the $c\bar{c}s\bar{s}$ tetraquark system and some experimental discovered charmonium-like state, such as $\chi_{c0}(3930)$, $X(4350)$, $X(4500)$, $X(4700)$ and $X(4274)$, coule be well explained. For the sake of completeness, we collect the relevant model parameters in Table~\ref{Tab:Para}.

In the QDCSM, the single-particle orbital wave functions in the ordinary quark cluster model are the left and right centered single Gaussian functions, which are,
\begin{eqnarray}\label{single}
\phi_\alpha(\boldsymbol {S_{i}})=\left(\frac{1}{\pi
b^2}\right)^{\frac{3}{4}}e^ {-\frac{(\boldsymbol {r_{\alpha}}-\frac{1}{2}\boldsymbol
{S_i})^2}{2b^2}},
 \nonumber\\
\phi_\beta(-\boldsymbol {S_{i}})=\left(\frac{1}{\pi
b^2}\right)^{\frac{3}{4}}e^ {-\frac{(\boldsymbol {r_{\beta}}+\frac{1}{2}\boldsymbol
{S_i})^2}{2b^2}} .
 \
\end{eqnarray}
The quark delocalization is realized by writing the single-particle orbital wave function as a
linear combination of the left and right Gaussians, which are,
\begin{eqnarray}
{\psi}_{\alpha}(\boldsymbol {S_{i}},\epsilon) &=&
\left({\phi}_{\alpha}(\boldsymbol{S_{i}})
+\epsilon{\phi}_{\alpha}(-\boldsymbol{S_{i}})\right)/N(\epsilon),
\nonumber \\
{\psi}_{\beta}(-\boldsymbol {S_{i}},\epsilon) &=&
\left({\phi}_{\beta}(-\boldsymbol{S_{i}})
+\epsilon{\phi}_{\beta}(\boldsymbol{S_{i}})\right)/N(\epsilon),
\nonumber \\
N(\epsilon)&=&\sqrt{1+\epsilon^2+2\epsilon e^{{-S}_i^2/4b^2}},
\end{eqnarray}
where $\epsilon(\boldsymbol{S}_i)$ is the delocalization parameter determined by the dynamics of the quark system rather than free parameters. In this way, the system can choose its most favorable configuration through its dynamics in a larger Hilbert space.

\subsection{The resonating group method}

\begin{figure}[htb]
\includegraphics[scale=0.65]{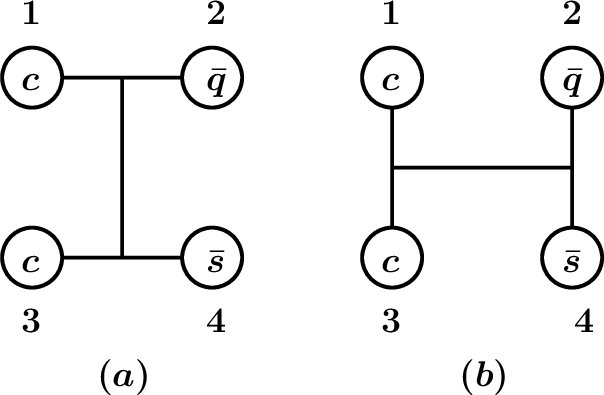}
\caption{The meson-meson configuration (diagram (a)) and diquark-antidiquark configuration (diagram (b)) in the $\tccs$ tetraquark system. }
\label{fig1}
\end{figure}

In the present work, the RGM is employed to carry out the dynamical calculation.  When dealing with the two-cluster system in this method, one can only consider the relative motion between the clusters, while the two clusters are frozen inside~\cite{Kamimura:1981oxj}. So the wave function of the $\tccs$ system can be constructed as,
\begin{eqnarray}\label{wave1}
  \psi_{4q} &=&  \mathcal{A}\left[\left[\psi_{A}(\boldsymbol{\rho}_A)\psi_{B}(\boldsymbol{\rho}_B)\right]^{[\sigma]IS}\otimes\chi_{L}(\textbf{R})\right]^{J},
\end{eqnarray}
where the symbol $\mathcal{A}$ is the antisymmetry operator, which is defined as
\begin{eqnarray}\label{wave2}
    \mathcal{A}&=& 1-P_{13}.
\end{eqnarray}
where the $P_{13}$ indicates the exchange of the particle positions with numbers $1$ and $3$ from the Fig.~\ref{fig1}. $[\sigma]=[222]$ gives the total color symmetry. The symbols $I$, $S$, $L$, and $J$ represent flavor, spin, orbit angular momentum, and total angular momentum of $\tccs$ system, respectively. $\psi_{A}$ and $\psi_{B}$ are the wave functions of the two-quark cluster, which are,
\begin{eqnarray}
 \psi_{A} &=& \left(\frac{1}{2\pi b^{2}}\right)^{3/4}  e^{-\boldsymbol{\rho_{A}}^{2}/(4b^2)} \eta_{I_{A}}S_{A}\chi_{A}^{c} ,\nonumber\\
  \psi_{B} &=& \left(\frac{1}{2\pi b^{2}}\right)^{3/4} e^{-\boldsymbol{\rho_{B}}^{2}/(4b^2)} \eta_{I_{B}}S_{B}\chi_{B}^{c} ,
\end{eqnarray}
where $\eta_I$, $S$, and $\chi$ represent the flavor, spin and internal color terms of the cluster wave functions, respectively. According to Fig.~\ref{fig1}, we adopt different Jacobi coordinates for different diagrams. As for the meson-meson configuration in Fig.~\ref{fig1}-(a), the Jacobi coordinates are defined as,
\begin{eqnarray}
\label{wave6}
\boldsymbol{\rho_{A}}&=&{\br_{q_1}-\br_{\bar{q}_2}},  \ \ \ \ \boldsymbol{\rho_{B}}={\br_{q_3}-\br_{\bar{q}_4}},\nonumber \\
\boldsymbol{R_{A}}&=& \frac{m_{1}{\br_{q_1}}+m_{2} {\br_{\bar{q}_2}}}{m_{1}+m_{2}},\nonumber \\
 \boldsymbol{R_{B}}&=& \frac{m_{3}{\br_{q_3}}+m_{4}{\br_{\bar{q}_4}}}{m_{3}+m_{4}},\nonumber \\
\boldsymbol{R}&=&\boldsymbol{R_{A}-R_{B}}, \nonumber \\
\boldsymbol{R_{c}}&=&\frac{m_{1}{\br_{q_1}}+m_{2}{\br_{\bar{q}_2}}+m_{3}{\br_{q_3}}+m_{4}{\br_{\bar{q}_4}}}{m_{1}+m_{2}+m_{3}+m_{4}}.
\end{eqnarray}
where the subscript $q/\bar{q}$ indicates the quark or antiquark particle, while the number indicates the quark position in Fig.~\ref{fig1}-(a). As for the diquark-antidiquark configuration as shown in Fig.~\ref{fig1}-(b), the relevant Jacobi coordinates can be obtained by interchanging $\br_{q_3}$ with $\br_{\bar{q}_2}$ in Eq.~(\ref{wave6}).

Form the variational principle, after variation with respect to the relative motion wave function $\chi\boldsymbol(R)=\sum_{L}\chi_{L}\boldsymbol(R)$, one obtains the RGM equation, which is,
\begin{eqnarray}\label{wave3}
  \int H\left(\boldsymbol{R, R^{\prime}}\right)\chi\left(\boldsymbol{R^{\prime}}\right)d\boldsymbol R^{\prime} =E\nonumber \\  \int N\left(\boldsymbol{R, R^{\prime}}\right) \chi\left(\boldsymbol{R^{\prime}}\right)d\boldsymbol R^{\prime},
\end{eqnarray}
where $H(\boldsymbol{R, R^{\prime}})$ and $N(\boldsymbol{R, R^{\prime}})$    are Hamiltonian and norm kernels, respectively. The eigenenergy $E$ and the wave functions can be obtained by solving the RGM equation. In the present estimation, the function $\chi (\boldsymbol{R})$ can be expanded by gaussian bases,  which is
\begin{eqnarray}\label{wave5}
\chi\boldsymbol{(R)}&=&\frac{1}{\sqrt{4\pi}}\sum_{L}\left(\frac{1}{\pi b^2}\right)^{3/4}\sum_{i}^{n}C_{i,L} \nonumber\\
&&\times\int e^{-\frac{1}{2}\boldsymbol(R-S_{i})^{2}/b^2} Y^L\left(\hat{\boldsymbol{S}_{i}}\right)d\hat{\boldsymbol{S}_{i}},
\end{eqnarray}
where $C_{i,L}$ is the expansion coefficient, and $n$ is the number of gaussian bases, which is determined by the stability of the results. $\boldsymbol{S}_{i}$ is the separation of two reference centers. $\boldsymbol{R}$ is the dynamic coordinate defined in Eq.~(\ref{wave6}). After including the motion of the center of mass, i.e., 
\begin{equation}
  \phi_{C}(\boldsymbol{R_{c}})=\left(\frac{4}{\pi b^2}\right)^{3/4}\mathrm{e}^{\frac{-2 \boldsymbol{R_{c}}^{2}}{b^{2}}},
\end{equation}
one can rewrite Eq.~(\ref{wave1}) as,
\begin{eqnarray}\label{wave4}
\psi_{4q}&=&\mathcal{A} \sum_{i,L}C_{i, L}\int \frac{d \hat{\boldsymbol{S_{i}}}}{\sqrt{4 \pi}} \prod_{\alpha=1}^{2}\phi_{\alpha}\left(\boldsymbol{S_{i}}\right)\prod_{\alpha=3}^{4}\phi_{\beta}\left(\boldsymbol{-S_{i}}\right) \nonumber \\
&&\times \left[\left[\eta_{I_{A}S_{A}}\eta_{I_{B}S_{B}}\right]^{IS}Y^{L}(\hat{\boldsymbol{S_{i}}})\right]^{J}\left[\chi_{A}^{c}\chi_{B}^{c}\right]^{[\sigma]} ,
\end{eqnarray}
where $\phi_{\alpha}(\boldsymbol{S_{i}})$ and $\phi_{\beta}(\boldsymbol{-S_{i}})$ are the single-particle orbital wave functions with different reference centers, whose specific expressions have been presented in Eq.~(\ref{single}).

With the reformulated ansatz as shown in Eq.~(\ref{wave4}), the RGM equation becomes an algebraic eigenvalue equation, which is,
\begin{eqnarray}
  \sum_{j,L}C_{J,L}H_{i,j}^{L,L^{\prime}} &=& E\sum_{j}C_{j,L^{\prime}}N_{i,j}^{L^{\prime}},
\end{eqnarray}
where $N_{i,j}^{L^{\prime}}$ and $H_{i,j}^{L,L^{\prime}}$ are the  overlap of the wave functions and the matrix elements of the Hamiltonian, respectively. By solving the generalized eigenvalue problem, we can obtain the energies of the tetraquark  systems $E$ and the corresponding expansion coefficients $C_{j,L}$. Finally, the relative motion wave function between two clusters can be obtained by substituting the $C_{j,L}$ into Eq.~(\ref{wave5}). As for the flavor, spin and color wave functions of the tetraquark system, they are constructed in a two step way. One can first construct the wave functions for the two clusters, and then coupling the wave functions of two clusters to form the wave function of tetraquark system. The details of the flavor, spin and color wave functions of tetraquark system are collected in the Appendix \ref{Sec:App}.

\begin{table*}[t]
\begin{center}
\caption{\label{channels} The relevant channels for all possible $J^P$ quantum numbers.}
\renewcommand\arraystretch{1.5}
\begin{tabular}{p{1.5cm}<\centering p{2.cm}<\centering p{1.2cm}<\centering p{0.2cm}<\centering p{1.5cm}<\centering p{2.cm}<\centering p{1.2cm}<\centering p{0.2cm}<\centering p{1.5cm}<\centering p{2.cm}<\centering p{1.2cm}<\centering p{1.5cm}<\centering}
\toprule[1pt]
\multicolumn{3}{c}{$J^{P}=0^{+}$} &  &\multicolumn{3}{c}{$J^{P}=1^{+}$} &  &\multicolumn{3}{c}{$J^{P}=2^{+}$} \\\cmidrule[1pt]{1-3} \cmidrule[1pt]{5-7} \cmidrule[1pt]{9-11}
\multirow{2}{*}{index}    &$F^{i}; S^{j}_{s}; \chi^{c}_{k}$   &\multirow{2}{*}{channels}  &     & \multirow{2}{*}{index}   &$F^{i}; S^{j}_{s}; \chi^{c}_{k}$   &\multirow{2}{*}{channels}  & & \multirow{2}{*}{index}   &$F^{i}; S^{j}_{s}; \chi^{c}_{k}$   &\multirow{2}{*}{channels}       \\
\multicolumn{1}{c}{} &[i;j;k] & \multicolumn{1}{c}{}  & &\multicolumn{1}{c}{} &[i;j;k] &\multicolumn{1}{c}{}  & &\multicolumn{1}{c}{} &[i;j;k] &\multicolumn{1}{c}{}   \\ \midrule[1pt]
1 & [1,1,1] &$D^{0}D_{s}^{+}$                       &  &1 & [1,3,1] &$D^{0}D_{s}^{*+}$                    &     &1 &[1,6,1] &$D^{*}D_{s}^{*+}$      \\
2 & [1,2,1] &$D^{*}D_{s}^{*+}$                      &  &2 & [1,4,1] &$D^{*}D_{s}^{+}$                     &     &2 &[2,6,4] &$(cc)(\bar{q}\bar{s})$       \\
3 & [2,1,3] &$(cc)(\bar{q}\bar{s})$                 &  &3 & [1,5,1] &$D^{*}D_{s}^{*+}$                    &     &\multicolumn{3}{c}{}         \\
4 & [2,2,4] &$(cc)(\bar{q}\bar{s})$                 &  &4 & [2,3,3] &$(cc)(\bar{q}\bar{s})$               &     &\multicolumn{3}{c}{}       \\
\multicolumn{3}{c}{}                    &  &5 & [2,4,4] &$(cc)(\bar{q}\bar{s})$   &      &\multicolumn{3}{c}{}       \\
\multicolumn{3}{c}{}                    &  &6 & [2,5,4] &$(cc)(\bar{q}\bar{s})$   &     &\multicolumn{3}{c}{}       \\
\bottomrule[1pt]
\end{tabular}
\end{center}
\end{table*}

\section{Numerical RESULTS AND DISCUSSIONS}{\label{results}}

In this work, only the low-lying $S-$wave $\tccs$ tetraquark state are considered and the spin of the tetraquark system can be 0, 1, and 2. Thus, the spin parity of $\tccs$ tetraquark states can be $0^+$, $1^+$ and $2^+$, respectively. Moreover, in the present estimations, both the meson-meson and diquark-antidiquark configurations are considered. In general, there are two types of color structures  for the meson-meson configuration, which are color singlet-singlet $( \textbf{1}_{c}\otimes\textbf{1}_{c})$ and the color octet-octet $(\textbf{8}_{c}\otimes \textbf{8}_{c})$. The later color structure have been taken into account by introducing the color screening effects in the model , thus, we only consider the color singlet-singlet structures in the present estimations. A for the diquark-antidiquark configuration, both the antitriplet-triplet $(\bar{\textbf{3}}_{c} \otimes \textbf{3}_{c})$ and sextet-antisextet $(\textbf{6}_{c}\otimes\bar{\textbf{6}}_{c})$ structure are taken into account. All the relevant  channels for all possible quantum numbers are listed in Table~\ref{channels}, where $F^{i}; S^{j}_{s}; \chi^{c}_{k}$ shows the necessary basis combinations in flavor ($F^{i}$), spin ($S^{j}_{s}$) and color ($\chi^{c}_{k}$) degrees of freedom.

\begin{table*}[!htb]
\begin{center}
\caption{\label{0bound} The low-lying eigenenergies (in unit of MeV) of $\tccs$ tetraquark states with $J^{P}=0^{+}$. }
\renewcommand\arraystretch{1.5}
\begin{tabular}{p{1.2cm}<\centering p{1.5cm}<\centering p{1.75cm}<\centering p{1.cm}<\centering p{1.cm}<\centering p{1.cm}<\centering p{0.5cm}<\centering p{1.cm}<\centering p{1.cm}<\centering p{0.5cm}<\centering p{0.5cm}<\centering p{1.cm}<\centering p{1.cm}<\centering p{1.2cm}<\centering p{1.5cm}<\centering }
\toprule[1pt]
\multirow{2}{*}{Index}  & \multirow{2}{*}{Channel}   & \multirow{2}{*}{Threshold}    &\multicolumn{3}{c}{QDCSM1} &  &\multicolumn{3}{c}{QDCSM2} &  & \multicolumn{3}{c}{QDCSM3} \\
\cmidrule[1pt]{4-6} \cmidrule[1pt]{8-10} \cmidrule[1pt]{12-14}
~~~  &  &  &$E_{sc}$   &$E_{cc}$  &$E_{mix}$ & &$E_{sc}$  &$E_{cc}$ &$E_{mix}$  & &$E_{sc}$  &$E_{cc}$ &$E_{mix}$\\
\midrule[1pt]
 ~~~1     &$D^{0}D_{s}^{+}$           &3833   &3836.3 &3836.2 &3836.2 &&3836.3 &3836.3 &3836.2   &&3836.2 &3836.2 &3836.2    \\
 ~~~2     &$D^{*}D_{s}^{*+}$          &4119   &4119.7 &       &       &&4120.9 &       &         &&4121.2 &     &        \\
 ~~~3     &$(cc)(\bar{q}\bar{s})$     &       &4589.3 &4299.8 &       &&4585.1 &4291.8 &         &&4574.7 &4277.9 &         \\
 ~~~4     &$(cc)(\bar{q}\bar{s})$     &       &4321.3 &       &       &&4316.5 &       &         &&4308.0 &               \\
\midrule[1pt]
\end{tabular}
\end{center}
\end{table*}

\begin{table*}[!htb]
\begin{center}
\caption{\label{1bound} The same as Table~\ref{0bound} but for the tetraquark states with $J^{P}=1^{+}$. }
\renewcommand\arraystretch{1.5}
\begin{tabular}{p{1.2cm}<\centering p{1.5cm}<\centering p{1.75cm}<\centering p{1.cm}<\centering p{1.cm}<\centering p{1.cm}<\centering p{0.5cm}<\centering p{1.cm}<\centering p{1.cm}<\centering p{0.5cm}<\centering p{0.5cm}<\centering p{1.cm}<\centering p{1.cm}<\centering p{1.2cm}<\centering p{1.5cm}<\centering }
\toprule[1pt]
\multirow{2}{*}{Index}  & \multirow{2}{*}{Channel}   & \multirow{2}{*}{Threshold}    &\multicolumn{3}{c}{QDCSM1} &  &\multicolumn{3}{c}{QDCSM2} &  & \multicolumn{3}{c}{QDCSM3} \\
\cmidrule[1pt]{4-6} \cmidrule[1pt]{8-10} \cmidrule[1pt]{12-14}
~~~  &  &   &$E_{sc}$   &$E_{cc}$  &$E_{mix}$ & &$E_{sc}$  &$E_{cc}$ &$E_{mix}$  & &$E_{sc}$  &$E_{cc}$ &$E_{mix}$\\ \midrule[1pt]
 ~~~1     &$D^{0}D_{s}^{*+}$         &3977   &3978.2 &3977.1 &3971.1 &&3978.2 &3977.7 &3973.8   &&3978.2 &3978.1 &3974.8    \\
 ~~~2     &$D^{*}D_{s}^{+}$          &3975   &3978.0 &       &       &&3978.1 &       &         &&3978.2 &       &        \\
 ~~~3     &$D^{*}D_{s}^{*+}$         &4119   &4110.8 &       &       &&4117.2 &       &         &&4118.1 &       &         \\
 ~~~4     &$(cc)(\bar{q}\bar{s})$    &       &4544.2 &4128.2 &       &&4535.4 &4127.2 &         &&4518.9 &4124.1             \\
 ~~~5     &$(cc)(\bar{q}\bar{s})$    &       &4132.7 &       &       &&4132.5 &       &         &&4130.7 &                    \\
 ~~~6     &$(cc)(\bar{q}\bar{s})$    &       &4337.5 &       &       &&4334.1 &       &         &&4327.8 &                    \\
\bottomrule[1pt]
\end{tabular}
\end{center}
\end{table*}

\begin{table*}[!htb]
\begin{center}
\caption{\label{2bound} The same as Table~\ref{0bound} but for the tetraquark states with $J^{P}=2^{+}$. }
\renewcommand\arraystretch{1.5}
\begin{tabular}{p{1.2cm}<\centering p{1.5cm}<\centering p{1.75cm}<\centering p{1.cm}<\centering p{1.cm}<\centering p{1.cm}<\centering p{0.5cm}<\centering p{1.cm}<\centering p{1.cm}<\centering p{0.5cm}<\centering p{0.5cm}<\centering p{1.cm}<\centering p{1.cm}<\centering p{1.2cm}<\centering p{1.5cm}<\centering }
\toprule[1pt]
\multirow{2}{*}{Index}  & \multirow{2}{*}{Channel}   & \multirow{2}{*}{Threshold}  &\multicolumn{3}{c}{QDCSM1} &  &\multicolumn{3}{c}{QDCSM2} &  & \multicolumn{3}{c}{QDCSM3} \\
\cmidrule[1pt]{4-6} \cmidrule[1pt]{8-10} \cmidrule[1pt]{12-14}
&  &    &$E_{sc}$   &$E_{cc}$  &$E_{mix}$ & &$E_{sc}$  &$E_{cc}$ &$E_{mix}$  & &$E_{sc}$  &$E_{cc}$ &$E_{mix}$\\\midrule[1pt]
 ~~~1     &$D^{*}D_{s}^{*+}$         &4119   &4122.0 &--- &4121.5 &&4122.2 &--- &4122.1   &&4122.3 &---&4122.2    \\
 ~~~2     &$(cc)(\bar{q}\bar{s})$    &       &4367.1 &--- &       &&4366.3 &--- &         &&4364.1 &---                  \\
\bottomrule[1pt]
\end{tabular}
\end{center}
\end{table*}

\subsection{Bound State}
With the above preparations, the low-lying $S-$wave $\tccs$ tetraquark states are systematically explored herein. In Tables~\ref{0bound}-~\ref{2bound}, we collect the estimated eigenenergies of the  $\tccs$ tetraquark states with different $J^P$ quantum numbers. In those tables, the index of the first column represents the symbols of each channel and in the second and third columns we list all the involved  channels and the corresponding theoretical threshold, respectively. Moreover, $E_{sc}$ is the eigenenergy obtained in the single channel estimations, $E_{cc}$ and $E_{mix}$ are the eigenenergies estimated by considering the coupled channel effects in each kind of configuration, and in both configurations, respectively.

Additionally, we define the binding energy $E_{b}$ of the $\tccs$ tetraquark states as $E_{bi}=E_{i}-E_{4}(\infty)$ to identify whether or not the tetraquark states are stable against the strong interactions, where $E_{4}(\infty)$ is the lowest possible threshold of the two meson structure estimated in the QDCSM. and $i$ represents the different situation of channel coupling. Such a subtraction procedure can greatly reduce the influence of the model parameters on the binding energies. If $E_{b} > 0$, the tetraqaurk systems can fall apart into two mesons via the strong interactions. If $E_{b}< 0$, the strong decay into two mesons is forbidden kinemetically and therefore the decay can only occur via either the weak or electromagnetic interaction.

For the $\tccs$ tetraquark system with $J^{P}=0^{+}$, there are two channels in the meson-configuration and two channels in the diquark-antidiquark configuration. The estimated eigenenergies of $\tccs$ state with $J^P=0^+$ are listed in Table~\ref{0bound}. The theoretical thresholds of the meson-meson channels are also presented for comparison. With the parameters in QDCSM1, the single channel estimations in the meson-meson configuration find that the eigenenergies are all above the corresponding threshold, which indicate that the single channel estimations do not support the existence of the bound states. In addition, when considering the coupled channels effects in the meson-meson configurations, we find the estimated eigenenergy is 3836.2 MeV, which is above the threshold of $D^0 D_s^+$. The lowest eigenenergy obtained by coupled channel estimations in the meson-meson configuration is very close to the one of the single channel estimations in the $D^0 D_s^+$ channel, which indicates that the coupled channel effect in the meson-meson configuration is rather weak. As for the diquark-antidiquark configuration, both the single channel estimations and the coupled channel estimations indicate that the eigenenergies are above the threshold of $D^0 D_s^+$. Different from the meson-meson configuration, we find the eigenenergy obtained from the coupled channel estimation  is at least 20 MeV below the lowest one of the single channel estimation, which indicate the coupled channels effect in the diquark-antidiquark configuration is much strong. Moreover, we extend the coupled channel effect in both configurations, and the eigenenergy is estimated to be 3836.2 MeV, which is still above the threshold of $D^0 D_s^+$. The results estimated with the parameters in QDCSM2 and QDCSM3 are very similar with those obtained with the parameter in QDCSM1 and no stable tetraquark state is found.

For the $\tccs$ tetraquark system with $J^{P}=1^{+}$, there are six channels, including three channels in the meson-meson configuration and three channels in the diquark-antidiquark configuration. From Table~\ref{1bound}, the estimated results of three sets of model parameters are almost identical. When considering the single channel estimations in the meson-meson configuration, we find that the estimated eigenenergy of $D^{0}D_{s}^{*+}$ and $D^{*}D_{s}^{+}$ channels are above the theoretical threshold of the corresponding physical channels, which indicates that these channels are scattering channels in single channel calculations. However, a bound state in the $D^{*}D_{s}^{*+}$ channel with the bound energy about $1\sim10$ MeV is obtained with all three sets of model parameters. Besides, by the coupling channels with the meson-meson configuration, the estimated eigenenergy is slightly above the lowest theoretical threshold of the $D^{*}D_{s}^{+}$, which show that the effect of couple channels in the meson-meson configuration is rather weak. For the diquark-antidiquark configuration, the estimated eigenenergies obtained for the single-channel and channel-coupled estimations are above the theoretical threshold of the lowest channel $D^{*}D_{s}^{+}$. Nevertheless, when the channel coupling between the two configuration are taken into account, a shallow bound state is detected, although the magnitude of the bound energy is slightly different with different sets of the model parameters.

\begin{table*}[t]
\begin{center}
\caption{\label{part1} Contributions of each terms in Hamiltonian to the energy of the $D^{0}D_{s}^{*+}$ bound state with $J^{P}=1^{+}$ in unit of MeV. $E_{M}$ stands for the sum of two mesons threshold. Our estimations indicate the contributions of $\eta$ meson exchange potential are all less than $0.05$ MeV in different sets of model parameters. Thus, the contributions from $\eta$ meson exchange are not presented.}
\renewcommand\arraystretch{1.75}
\begin{tabular}{p{1.4cm}<\centering p{1.15cm}<\centering p{1.15cm}<\centering p{1.15cm}<\centering p{1.15cm}<\centering p{0.1cm}<\centering %
    p{1.15cm}<\centering p{1.15cm}<\centering p{1.15cm}<\centering p{1.15cm}<\centering p{0.1cm}<\centering %
    p{1.15cm}<\centering p{1.15cm}<\centering p{1.15cm}<\centering p{1.15cm}<\centering}
\toprule[1pt]
       & \multicolumn{4}{c}{QDCSM1} &&\multicolumn{4}{c}{QDCSM2} &&\multicolumn{4}{c}{QDCSM3}\\
        \cmidrule[1pt]{2-5}   \cmidrule[1pt]{7-10}  \cmidrule[1pt]{12-15}
        &$H_{T}$           &$V_{\mathrm{CON}}$     &$V_{\mathrm{OGE}}$      &$V_{K}$  &
        &$H_{T}$           &$V_{\mathrm{CON}}$     &$V_{\mathrm{OGE}}$      &$V_{K}$  &
        &$H_{T}$           &$V_{\mathrm{CON}}$     &$V_{\mathrm{OGE}}$      &$V_{K}$\\
              \midrule[1pt]
   $E_{sc}$             &1081.3      &-901.7     &-506.6         &$\sim0.0$  &
                        &1011.2      &-783.9     &-554.2         &$\sim0.0$  &
                        &917.9       &-615.9     &-628.8         &$\sim0.0$  \\
   $E_{cc}$             &1073.9      &-895.9     &-505.8         &-0.1       &
                        &1008.8      &-782.5     &-553.5         &-0.1       &
                        &917.1       &-615.5     &-628.5         &$\sim0.0$  \\
   $E_{mix}$            &1049.0      &-820.4     &-558.1         &-4.4       &
                        &998.4       &-752.4     &-573.7         &-3.5       &
                        &915.3       &-609.8     &-635.4         &-0.3       \\
   $E_{M}$              &1079.6      &-903.3     &-506.1         &$\sim0.0$  &
                        &1008.7      &-784.7     &-553.8         &$\sim0.0$  &
                        &915.0       &-616.3     &-628.5         &$\sim0.0$  \\
   \midrule[1pt]
   $\Delta E_{sc}$      &1.7         &1.6        &-0.5           &$\sim0.0$  &
                        &2.5         &0.8        &0.4            &$\sim0.0$  &
                        &2.9         &0.4        &-0.3           &$\sim0.0$  \\
   $\Delta E_{cc}$      &-5.7        &7.4        &0.3            &-0.1       &
                        &0.1         &2.2        &-0.3           &-0.1       &
                        &2.1         &0.8        &0.0            &$\sim0.0$  \\
   $\Delta E_{mix}$     &-30.6       &82.9       &-52.0          &-4.4       &
                        &-10.3       &32.3       &-19.9          &-3.5       &
                        &0.3         &5.5        &-7.2           &-0.3       \\
\bottomrule[1pt]
\end{tabular}
\end{center}
\end{table*}

In view of the above conclusions, we estimate the average values of each terms in the Hamiltonian to examine how a shallow $D^{*}D_{s}^{+}$ bound state with $J^{P}=1^{+}$ is created. In Table~\ref{part1}, we present the  contributions of each interaction by considering the single channel and coupled channel calculations. In addition, the average values of each interaction of two conventional $D^{*}$and $D_{s}^{+}$ mesons  without interactions, i.e., the distance between the two mesons are large enough, are also listed in the table for comparison. From the Table, one finds that the magnitude of the average values of each terms for different sets of model parameter are very similar. Here, we define $\Delta E_{sc}= E_{sc}- E_M$, $\Delta E_{cc} =E_{cc}-E_M$ and $\Delta E_{mix} =E_{mix}-E_M$. From our estimations, we find the contributions of the confinement potential to $\Delta E_{sc}$, $\Delta E_{cc}$ and $\Delta{E}_{mix}$ are positive, which indicate the confinement potential hinders the $D^{*}$ and $D_{s}^{+}$ subclusters from forming a bound states. For the kinetic energy term, with more physical channels taking into consideration, the properties of kinetic energy basically transforms gradually from repulsion towards very strong attraction, whereas the similar conclusions can be drawn for the one-gluon-exchange interaction. In addition, in the meson exchange interactions, the meson exchange potential contributions to $\Delta E_{sc}$, $\Delta E_{cc}$ and $\Delta{E}_{mix}$ are negligible, in particularly, the contributions from $\eta$ meson exchange potential are all less than 0.05 MeV, which are not listed in the table. According to the above analysis, one can find that the kinetic energy term and the one-gluon-exchange  potential have deep attractions in the channel coupling calculations with both the meson-meson and diquark-antidiquark configurations, However, the confinement potential displays as  repulsive interaction, which weaken the overall attraction. Such a phenomenon illustrates the very delicate competition between the kinetic energy and the interaction potential from various sources in the Hamiltonian.

For the $\tccs$ tetraquark system with $J^{P}=2^{+}$, only one physics channel in the meson-meson configuration and one channel in the diquark-antidiquark configuration exists. From Table~\ref{2bound}, one can find the eigenenergies obtained from the single channel estimation is higher than the physical meson-meson channel. After considering the coupled channel effect between the meson-meson and diquark-antidiquark configurations, the estimated eigenenergy is still above the threshold of $D^\ast D_s^{\ast +}$, which indicates that there is no bound state in the $\tccs$ tetraquark system with $J^{P}=2^{+}$.

  \begin{figure}
\includegraphics[scale=0.70]{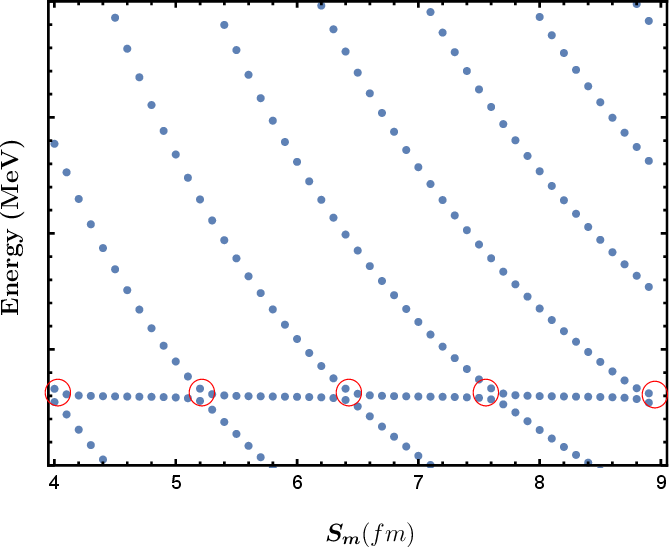}
\caption{A sketch diagram of the resonance shape in the real-scaling method.  }
\label{real_scal1}
\end{figure}

\subsection{Resonance States}
In the bound state estimations, we find one bound state with $J^P=1^+$ while there is no bound state in the $J^{P}=0^+$ and $J^{P}=2^+$ systems. In the following, we will employ the real scaling method to explore the possible resonance states in the $\tccs$ tetraquark system. To determine whether these resonance states could be detected by the open channels, we perform a channel coupling estimation by including all the meson-meson and diquark-antidiquark channels in the estimations.

 The real scaling method is developed to identify the genuine resonances from the states with discrete energies with finite volume~\cite{real_Method} . In this method, a factor $\textbf{S}_{m}$, which is the distance between two clusters, is adopted to scale the finite volume. So with the increase of the distance between two clusters, the continuum state will fall off toward its threshold, the energy of the bound state remains unchanged, while a resonance state will tend to be stable. If the energy of a scattering state is far away from the one of the resonance, the coupling between the resonance and the scattering states is rather weak, and the energy of the resonance is almost stable. When the energy of the scattering state approaches the one of the resonance due to the increasing of $\textbf{S}_{m}$, the coupling will become strong, and if $\textbf{S}_{m}$ increases further, the energy gap between the resonance and scattering states will increase and the coupling will become weak again. In this way, an avoided crossing structure appears. This is a general feature of two interacting energy levels. Because of the continuum nature of the scattering states, the avoided crossing structure will show up repeatedly with the increasing of  $\textbf{S}_{m}$ as shown in Fig.~\ref{real_scal1}  and the resonance line corresponds to the energy of the resonance state.  In addition, from the slopes of resonance and scattering states, the decay width can be estimated by,
\begin{eqnarray}\label{width}
  \Gamma &=& 4|V_{min}(S)|\frac{\sqrt{|k_{r}||k_{c}|} }{|k_{r}-k_{c}|}
\end{eqnarray}
 where $k_{r}$ and $k_{c}$ are the slopes of the resonance and scattering states, respectively. While, $V_{min}(S)$ is the minimal energy difference between the resonance and the scattering state at avoided crossing point. This method has been successfully applied to investigate the pentaquark ~\cite{Hiyama:2005cf,Hiyama:2018ukv}, the dibaryon ~\cite{Xia:2021tof}, and the tetraquark systems~\cite{Jin:2020jfc,Liu:2021xje,Liu:2022vyy}.

\begin{figure*}[htb]
\includegraphics[scale=2]{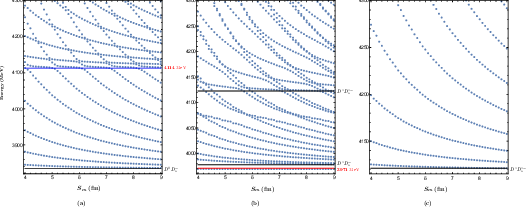}
\caption{The stabilization plots of the energies of the $\tccs$ tetraquark systems. \label{Fig:Resonance}}
\end{figure*}

  In the present work, we expand the spacial wave function with a set of gaussians with differences $\textbf{S}_{m}$, $(m=1,2,3,\ldots,n)$ and the distance with the relative motion of two clusters can be scaled. So we calculate the energy eigenvalues of the $\tccs$ tetraquark system by taking the value of the largest distance $(S_{m})$ between two clusters from 4.0 to 9.0 fm to check if there is any resonance state. Here, we take the results of the QDCSM1 as examples, which are shown in Fig.~\ref{Fig:Resonance} with different $J^P$ quantum numbers. For the $\tccs$ tetraquark system with $J^{P}=0^+$ as shown in Fig.~\ref{Fig:Resonance}-(a), one can note that the lower black horizontal line corresponds to the physical threshold of $D_{s}^{+}D^{0}$, while the upper blue horizontal line with the energy to be about 4114 MeV, locates below the threshold of $D^{*}D_{s}^{+*}$, which corresponds to a resonance state since the resonance behavior appearing in the Fig.~\ref{Fig:Resonance}-(a) as the finite space is constantly expanding. Moreover, the resonance state is estimated by considering the full channel coupling, and the present result indicates that its main ingredient is $D^{*}D_{s}^{+*}$. In other words, the effect of the channel coupling push the energy of the physical channel $D^{*}D_{s}^{+*}$ a bit below its threshold. In addition, the width of this resonance state is estimated to be about $14.3$ MeV according to Eq.~(\ref{width}).

For the $\tccs$ tetraquark system with $J^{P}=1^+$ as shown in Fig.~\ref{Fig:Resonance}-(b), it is obvious that the lowest red horizontal line locates at the energy of $3971$ MeV, which is below the threshold of the $D^{0}D_{s}^{+*}$, and this represents the bound states of $\tccs$ tetraquark system with $J^{P}=1^+$. This conclusion is consistent with the estimations in the last subsection. Moreover, two additional horizontal lines are also presented, which stand for the threshold of $D^{*}D_{s}^{+}$ and $D^{*}D_{s}^{*+}$, respectively. The present estimations indicate that there is no resonance state in the $\tccs$ tetraquark system with $J^{P}=1^+$, and the bound state in the $D^{*}D_{s}^{*+}$ channel becomes the scattering state by the effect of the channel coupling. For the $\tccs$ tetraquark system with $J^{P}=2^+$ as shown in Fig.~\ref{Fig:Resonance}-(c), there is one horizontal line, which  represents the threshold of $D^{*}D_{s}^{*+}$. It is clearly to conclude that there are no bound or resonant states in the $\tccs$ tetraquark system with $J^{P}=2^+$.

In addition, we perform the same estimations for the $\tccs$ tetraquark system in the QDCSM2 and QDCSM3. The results are similar to those of QDCSM1. We summarize the results obtained from three sets of model parameters in Table~\ref{resonance}.  By taking the coupled channel effects into consideration, we find one resonance state with a mass $4113\sim4114$ MeV for the $\tccs$ tetraquark system with $J^{P}=0^+$. The dominant component of the resonance state is $D^{*}D_{s}^{*+}$ with the percentage of this component to be about $80\%$. Moreover, the decay width of this resonance state is predicted to be $14.3\sim 16.1$ MeV. For the $J^{P}=1^+$ system, there is a bound state with energy range  $(3971.1\sim3974.8)$ MeV and no resonance state is obtained. For the $\tccs$ tetraquark system with $J^P=2^+$, no resonance or bound state is obtained by the channel coupling estimations.

\begin{table}[htb]
\begin{center}
\caption{The energies and widths of the $\tccs$ tertraquark states. \label{resonance} }
\renewcommand\arraystretch{1.2}
\begin{tabular}{p{1.2cm}<\centering  p{0.8cm}<\centering p{1.9cm}<\centering p{1.9cm}<\centering p{1.9cm}<\centering  }
\toprule[1pt]
\multicolumn{2}{c}{State} & \multicolumn{3}{c}{Parameter Sets}\\
\midrule[1pt]
& $J^P$ & QDCSM1 & QDCSM2 & QDCSM3  \\
\midrule[1pt]
 Bound & $1^+$ &  3971.1 & 3973.8 & 3974.8\\
 Resonance & $0^+$ & $4114/14.3$ & $4144/15.8$ & $4143/16.1$ \\
\bottomrule[1pt]
\end{tabular}
\end{center}
\end{table}


\section{Summary}\label{Sum}
In the present work, the $\tccs$ tetraquark system with the quantum number $J^{P}=0^+, 1^+, 2^+$ are systemically investigated to search for the possible bound state and resonance state by using the RGM in the QDCSM framework. In the model, both meson-meson and diquark-antidiquark configurations are taken into account, and the single-channel and the coupled channel calculations are preformed to obtain the energy of the $\tccs$ tetraquark system. In addition, a stabilization calculation is carried out to seek for possible resonance states. Furthermore, to check whether the estimated results are parameter dependent, three different sets of model parameters are employed in the calculation and we find the qualitative results of three sets of model parameters for the $\tccs$ tetraquark system are very similar.

From the present estimations, we find that the coupled channel effects plays important role in the $\tccs$ tetraquark system. After taking the coupled channel effects into consideration, we predict one bound state with the energy to be $3971.1 \sim3974.8$ MeV and $J^{P}=1^+$. Moreover, one resonance state with $J^{P}=0^+$ is also obtained, the resonance mass and width are estimated to be $4113\sim4114$ MeV and $14.3\sim 16.1$ MeV, respectively. The predictions in the present work could be experimentally detected in the future by LHCb and Belle II. Additionally the theoretical and further experimental investigations for properties of the $\tccs$ tetraquark could pave the way for possible doubly and triply tetraquark states.

\acknowledgments{This work is supported partly by the National Natural Science Foundation of China under
Contract No. 12175037, No. 12335001, No. 11775118 and No. 11535005. This work is also supported by china Postdoctoral Science Foundation funded project No. 2021M690626, and  No. 1107020201.

\appendix

\section{The wave function of the open heavy charm tetraquark with strangeness \label{Sec:App}}

\subsection{The color wave function}
Plenty of color structures in multiquark systems will be available with respect to those of conventional hadrons such as $q\bar{q}$ mesons and $qqq$ baryons. In this appendix, we present how to construct the colorless wave function for a tetraquark system.

For the meson-meson configurations, the color wave functions of a $q\bar{q}$ cluster would be,
\begin{eqnarray}
\nonumber
C^{1}_{[111]} &=& \sqrt{\frac{1}{3}}\Big(r\bar{r}+g\bar{g}+b\bar{b}\Big), \\ \nonumber
C^{2}_{[21]} &=& r\bar{b}, \qquad C^{3}_{[21]} =  -r\bar{g},                    \\ \nonumber
C^{4}_{[21]} &=& g\bar{b}, \qquad C^{5}_{[21]} =  -b\bar{g},              \\ \nonumber
C^{6}_{[21]} &=& g\bar{r}, \qquad C^{7}_{[21]} =   b\bar{r},         \\ \nonumber
C^{8}_{[21]} &=&  \sqrt{\frac{1}{2}}\Big(r\bar{r}-g\bar{g}\Big),       \\
C^{9}_{[21]} &=&  \sqrt{\frac{1}{6}}\Big(-r\bar{r}-g\bar{g}+2b\bar{b}\Big),
\end{eqnarray}
where the subscript [111] and [21] stand for color-singlet ($\textbf{1}_{c}$) and color-octet ($\textbf{8}_{c}$), respectively. So, the $\mathrm{SU}(3)_{\mathrm{color}}$ wave functions of color-singlet (two color-singlet cluters, $\textbf{1}_{c}\otimes\textbf{1}_{c}$) and hidden-color (two color-octet clusters, $\textbf{8}_{c}\otimes\textbf{8}_{c}$) channels are given, respectively,
\begin{eqnarray}
\chi^{c}_{1} &=& C^{1}_{[111]}C^{1}_{[111]}, \nonumber\\
  \chi^{c}_{2} &=&\sqrt{\frac{1}{8}}\Big(C^{2}_{[21]}C^{7}_{[21]}-C^{4}_{[21]}C^{5}_{[21]}-C^{3}_{[21]}C^{6}_{[21]}\nonumber \\
               &&+C^{8}_{[21]}C^{8}_{[21]}-C^{6}_{[21]}C^{3}_{[21]}+C^{9}_{[21]}C^{9}_{[21]}\nonumber \\
               &&-C^{5}_{[21]}C^{4}_{[21]}+C^{7}_{[21]}C^{2}_{[21]}\Big).
\end{eqnarray}

For the diquark-antidiquark structure, the color wave functions of the diquark clusters are,
\begin{eqnarray}
\nonumber
  C^{1}_{[2]} &=& rr, \qquad C^{2}_{[2]} = \sqrt{\frac{1}{2}}\Big(rg+gr\Big), \\ \nonumber
  C^{3}_{[2]} &=& gg,  \qquad C^{4}_{[2]} = \sqrt{\frac{1}{2}}\Big(rb+br\Big),\\ \nonumber
  C^{5}_{[2]} &=& \sqrt{\frac{1}{2}}\Big(gb+bg\Big), \qquad C^{6}_{[2]} = bb, \\ \nonumber
  C^{7}_{[11]} &=& \sqrt{\frac{1}{2}}\Big(rg-gr\Big),\qquad   C^{8}_{[11]} = \sqrt{\frac{1}{2}}\Big(rb-br\Big), \\
    C^{9}_{[11]} &=&\sqrt{\frac{1}{2}}\Big(gb-bg\Big).
\end{eqnarray}
While the color wave functions of the antidiquark clusters can be writen as,
\begin{eqnarray}
\nonumber
  C^{1}_{[22]} &=& \bar{r}\bar{r}, \qquad C^{2}_{[22]} = -\sqrt{\frac{1}{2}}\Big(\bar{r}\bar{g}+\bar{g}\bar{r}\Big),\\  \nonumber
  C^{3}_{[22]} &=& \bar{g}\bar{g}, \qquad C^{4}_{[22]} = \sqrt{\frac{1}{2}}\Big(\bar{r}\bar{b}+\bar{b}\bar{r}\Big), \\  \nonumber
  C^{5}_{[22]} &=& -\sqrt{\frac{1}{2}}\Big(\bar{g}\bar{b}+\bar{b}\bar{g}\Big), \qquad C^{6}_{[22]} = \bar{b}\bar{b}, \\ \nonumber
  C^{7}_{[211]}&=& \sqrt{\frac{1}{2}}\Big(\bar{r}\bar{g}-\bar{g}\bar{r}\Big),\qquad C^{8}_{[211]} = -\sqrt{\frac{1}{2}}\Big(\bar{r}\bar{b}-\bar{b}\bar{r}\Big), \\
  C^{9}_{[211]} &=& \sqrt{\frac{1}{2}}\Big(\bar{g}\bar{b}-\bar{b}\bar{g}\Big).
\end{eqnarray}
The color-singlet wave functions of the diquark-antidiquark configuration can be the product of color sextet and antisextet clusters ($\textbf{6}_{c}\otimes\bar{\textbf{6}}_{c}$) or the product of color-triplet and antitriplet cluster ($\textbf{3}_{c}\otimes\bar{\textbf{3}}_{c}$), which read,
\begin{eqnarray}
\chi^{c}_{3} &= &\sqrt{\frac{1}{6}}\Big(C^{1}_{[2]}C^{1}_{[22]}-C^{2}_{[2]}C^{[2]}_{[22]}+C^{3}_{[2]}C^{3}_{[22]} \nonumber \\
              & &+C^{4}_{[2]}C^{4}_{[22]}-C^{5}_{[2]}C^{5}_{[22]}+C^{6}_{2}C^{6}_{22}\Big),\nonumber\\
  \chi^{c}_{4} &=&\sqrt{\frac{1}{3}}\Big(C^{7}_{[11]}C^{7}_{[211]}-C^{8}_{[11]}C^{8}_{[211]}+C^{9}_{[11]}C^{9}_{[211]}\Big).
\end{eqnarray}

\subsection{The flavor wave function}
For the flavor degree of freedom, the different coupling methods generate different flavor wave function. From the Table~\ref{fig1}, the $\tccs$ tetraquark flavor wave function can be categorized as $F^{i}_{m}$ and $F_{d}^{i}$, where the subscript $m$ and $d$ refer to meson-meson and the diquark-antidiquark configurations, respectively. Distinctive structures are gotten the quark coupling arrange.  For the meson-meson structure, the coupling orders can be accessed as,
\begin{eqnarray}
  F^{1}_{m}&=& (c\bar{q})-(c\bar{s}),
\end{eqnarray}
while for the diquark-antidiquark structure, the flavor wave function should be written as
\begin{eqnarray}
    F^{2}_{d}&=& (cc)-(\bar{q}\bar{s})
\end{eqnarray}

\subsection{The spin wave function}
The total spin $S$ of tetraquark states ranges from 0 to 2. All of them are considered.
The wave functions of two body clusters are,
\begin{eqnarray}
\nonumber \chi_{11}&=& \alpha\alpha,\\
\nonumber \chi_{10} &=& \sqrt{\frac{1}{2}}\Big(\alpha\beta+\beta\alpha\Big),\\
\nonumber \chi_{1-1} &=& \beta\beta ,\\
            \chi_{00} &=& \sqrt{\frac{1}{2}}\Big(\alpha\beta-\beta\alpha\Big).
\end{eqnarray}

Then, the total spin wave functions $S^{i}_{s}$ are obtained by considering the coupling of two subcluster spin wave functions with SU(2) algebra, and the total spin wave functions of four-quark states can be read as,
\begin{eqnarray}
\nonumber S^{1}_{0}&=&\chi_{00}\chi_{00},\\
\nonumber S^{2}_{0}&=&\sqrt{\frac{1}{3}}\Big(\chi_{11}\chi_{1-1}-\chi_{10}\chi_{10}+\chi_{1-1}\chi_{11}\Big),\\
\nonumber S^{3}_{1}&=&\chi_{00}\chi_{11},\\
\nonumber S^{4}_{1}&=&\chi_{11}\chi_{00},\\
\nonumber S^{5}_{1}&=&\sqrt{\frac{1}{2}}\Big(\chi_{11}\chi_{10}-\chi_{10}\chi_{11}\Big), \\
S^{6}_{2}&=&\chi_{11}\chi_{11}.
\end{eqnarray}

\end{document}